\documentclass[reqno,11pt]{amsart}
\usepackage[utf8]{inputenc}
\usepackage{graphicx}
\usepackage{slashed}
\usepackage{amscd}
\usepackage{amssymb}
\usepackage{esint}
\usepackage{enumitem}
\usepackage{dsfont}
\usepackage[mathscr]{eucal}
\usepackage{upgreek}
\usepackage[off]{auto-pst-pdf} 
\usepackage{pst-all}

\numberwithin{equation}{section}
\allowdisplaybreaks[1]

\title[Causal Fermion Systems: Classical Gravity and Beyond]{Causal Fermion Systems: \\ Classical Gravity and Beyond}
\author[F.\ Finster]{Felix Finster \\ \\ September 2021}
\address{Fakult\"at f\"ur Mathematik  Universit\"at Regensburg  D-93040 Regensburg  Germany}
\email{finster@ur.de}

\newtheorem{Def}{Definition}[section]

\newcommand{\Thanks}{\vspace*{.5em} \noindent \thanks}
\newcommand{\beq}{\begin{equation}}
\newcommand{\eeq}{\end{equation}}
\newcommand{\Proof}{\begin{proof}}
	\newcommand{\QED}{\end{proof} \noindent}

\newcommand{\la}{\langle}
\newcommand{\ra}{\rangle}

\newcommand{\Sl}{\mbox{$\prec \!\!$ \nolinebreak}}
\newcommand{\Sr}{\mbox{\nolinebreak $\succ$}}

\newcommand{\C}{\mathbb{C}}
\newcommand{\R}{\mathbb{R}}
\newcommand{\1}{\mbox{\rm 1 \hspace{-1.05 em} 1}}

\newcommand{\N}{\mathbb{N}}

\DeclareMathOperator{\tr}{tr}

\renewcommand{\L}{{\mathcal{L}}}
\newcommand{\Sact}{{\mathcal{S}}}

\newcommand{\Cisc}{C^\infty_{\text{\rm{sc}}}}

\newcommand{\Dir}{{\mathcal{D}}}
\renewcommand{\H}{\mathscr{H}}

\newcommand{\Lin}{\text{\rm{L}}}
\newcommand{\F}{{\mathscr{F}}}

\DeclareMathOperator{\supp}{supp}

\newcommand{\scrM}{\mycal M}
\newcommand{\scrN}{\mycal N}

\newcommand{\scrU}{{\mathscr{U}}}

\newcommand{\s}{\mathfrak{s}}

\newcommand{\bitem}{\begin{itemize}[leftmargin=2.5em]}
\newcommand{\eitem}{\end{itemize}}

\newcommand{\G}{\mathscr{G}}

\newcommand{\Mass}{{\mathfrak{M}}}

\newcommand{\A}{\myscr A}

\newcommand{\x}{\mathbf{x}}
\newcommand{\y}{\mathbf{y}}

\DeclareFontFamily{OT1}{rsfso}{}
\DeclareFontShape{OT1}{rsfso}{m}{n}{ <-7> rsfso5 <7-10> rsfso7 <10-> rsfso10}{}
\DeclareMathAlphabet{\mycal}{OT1}{rsfso}{m}{n}
\DeclareMathAlphabet{\myscr}{OT1}{rsfso}{m}{n}

\begin{document}
\maketitle

\begin{abstract}
In this short review, we explain how and in which sense the 
causal action principle for causal fermion systems gives rise to classical gravity and the Einstein equations.
Moreover, methods are presented for going beyond classical gravity, with applications
to a positive mass theorem for static causal fermion systems,
a connection between area change and matter flux and
the construction of a quantum state.
\end{abstract}

\tableofcontents

\section{Introduction} \label{secintro}
The theory of {\em{causal fermion systems}} is a recent approach to fundamental physics
(see a few basics in Section~\ref{secprelim}, the reviews~\cite{nrstg, review}, the 
textbooks~\cite{cfs, intro} or the website~\cite{cfsweblink}).
In this approach, spacetime and all objects therein are described by a measure~$\rho$
on a set~$\F$ of linear operators on a Hilbert space~$(\H, \la .|. \ra_\H)$. 
The physical equations are formulated via the so-called {\em{causal action principle}},
a nonlinear variational principle where an action~$\Sact$ is minimized under variations of the measure~$\rho$.

The purpose of this survey article is to report on the present status of the theory
with regards to the gravitational interaction. It has been shown that, in a well-defined limiting case,
the so-called {\em{continuum limit}},
the causal action principles gives classical gravity and the Einstein equations,
with the gravitational coupling constant determined by the regularization scale.
Without taking the continuum limit, the causal action principle describes a novel gravitational theory
which is presently under investigation. We here give a survey on different approaches
towards unraveling the structure and the properties of this gravitational theory.

The paper is structured as follows. After giving the necessary background on causal fermion systems (Section~\ref{secprelim}), we explain how and in which sense classical gravity and the Einstein equations
are obtained in the continuum limit (Section~\ref{secclass}).
In Section~\ref{secbeyond} we report on the approaches for going beyond classical gravity.
After a few comments on the intrinsic geometric structures of a causal fermion system (Section~\ref{seclqg}),
we introduce surface layer integrals as the main object for the subsequent analysis (Section~\ref{secsli}).
We finally report on a positive mass theorem for static causal fermion systems (Section~\ref{secpmt}), 
a relation between area change and matter flux (Section~\ref{secjacobson}) and
the construction of a quantum state (Section~\ref{secQFT}).

\section{A Brief Introduction to Causal Fermion Systems} \label{secprelim}
This section provides the necessary abstract background on causal fermion systems.

\subsection{Causal Fermion Systems and the Causal Action Principle}
We begin with the general definitions.
\begin{Def} \label{defcfs} (causal fermion systems) {\em{ 
Given a separable complex Hilbert space~$\H$ with scalar product~$\la .|. \ra_\H$
and a parameter~$n \in \N$ (the {\em{spin dimension}}), we let~$\F \subset \Lin(\H)$ be the set of all
symmetric operators on~$\H$ of finite rank, which (counting multiplicities) have
at most~$n$ positive and at most~$n$ negative eigenvalues. On~$\F$ we are given
a positive measure~$\rho$ (defined on a $\sigma$-algebra of subsets of~$\F$).
We refer to~$(\H, \F, \rho)$ as a {\em{causal fermion system}}.
}}
\end{Def} \noindent
A causal fermion system describes a spacetime together
with all structures and objects therein.
In order to single out the physically admissible
causal fermion systems, one must formulate physical equations. To this end, we impose that
the measure~$\rho$ should be a minimizer of the causal action principle,
which we now introduce. For any~$x, y \in \F$, the product~$x y$ is an operator of rank at most~$2n$. 
However, in general it is no longer symmetric because~$(xy)^* = yx$,
and this is different from~$xy$ unless~$x$ and~$y$ commute.
As a consequence, the eigenvalues of the operator~$xy$ are in general complex.
We denote the non-trivial eigenvalues counting algebraic multiplicities
by~$\lambda^{xy}_1, \ldots, \lambda^{xy}_{2n} \in \C$
(more specifically,
denoting the rank of~$xy$ by~$k \leq 2n$, we choose~$\lambda^{xy}_1, \ldots, \lambda^{xy}_{k}$ as all
the non-zero eigenvalues and set~$\lambda^{xy}_{k+1}, \ldots, \lambda^{xy}_{2n}=0$).
We introduce the Lagrangian and the causal action by
\begin{align}
\text{\em{Lagrangian:}} && \L(x,y) &= \frac{1}{4n} \sum_{i,j=1}^{2n} \Big( \big|\lambda^{xy}_i \big|
- \big|\lambda^{xy}_j \big| \Big)^2 \label{Lagrange} \\
\text{\em{causal action:}} && \Sact(\rho) &= \iint_{\F \times \F} \L(x,y)\: d\rho(x)\, d\rho(y) \:. \label{Sdef}
\end{align}
The {\em{causal action principle}} is to minimize~$\Sact$ by varying the measure~$\rho$
under the following constraints,
\begin{align}
\text{\em{volume constraint:}} && \rho(\F) = \text{const} \quad\;\; & \label{volconstraint} \\
\text{\em{trace constraint:}} && \int_\F \tr(x)\: d\rho(x) = \text{const}& \label{trconstraint} \\
\text{\em{boundedness constraint:}} && \iint_{\F \times \F} 
|xy|^2
\: d\rho(x)\, d\rho(y) &\leq C \:, \label{Tdef}
\end{align}
where~$C$ is a given parameter, $\tr$ denotes the trace of a linear operator on~$\H$, and
the absolute value of~$xy$ is the so-called spectral weight,
\[ 
|xy| := \sum_{j=1}^{2n} \big|\lambda^{xy}_j \big| \:. \]
This variational principle is mathematically well-posed if~$\H$ is finite-dimensional.
For the existence theory and the analysis of general properties of minimizing measures
we refer to~\cite{continuum, lagrange} or~\cite[Chapter~12]{intro}.
In the existence theory one varies in the class of regular Borel measures
(with respect to the topology on~$\Lin(\H)$ induced by the operator norm),
and the minimizing measure is again in this class. With this in mind, here we always assume that
\[ 
\text{$\rho$ is a regular Borel measure}\:. \]

\subsection{Spacetime and Causal Structure}
Let~$\rho$ be a {\em{minimizing}} measure. {\em{Spacetime}}
is defined as the support of this measure,
\[ 
M := \supp \rho \;\subset\; \F \:. \]
Thus the spacetime points are symmetric linear operators on~$\H$.
On~$M$ we consider the topology induced by~$\F$ (generated by the operator norm
on~$\Lin(\H)$). Moreover, the measure~$\rho|_M$ restricted to~$M$ gives a volume
measure on spacetime. This gives spacetime the structure of a {\em{topological measure space}}.

The operators in~$M$ contain a lot of information which, if interpreted correctly,
gives rise to spacetime structures like causal and metric structures, spinors
and interacting fields (for details see~\cite[Chapter~1]{cfs}).
All the resulting objects are {\em{inherent}} in the sense that we only use information
already encoded in the causal fermion system.
Here we restrict attention to those structures needed in what follows.
We begin with the following notion of causality:

\begin{Def} (causal structure) \label{def2} 
{\em{ For any~$x, y \in \F$, we again denote the non-trivial ei\-gen\-values of the operator product~$xy$
(again counting algebraic multiplicities) by~$\lambda^{xy}_1, \ldots, \lambda^{xy}_{2n}$.
The points~$x$ and~$y$ are
called {\em{spacelike}} separated if all the~$\lambda^{xy}_j$ have the same absolute value.
They are said to be {\em{timelike}} separated if the~$\lambda^{xy}_j$ are all real and do not all 
have the same absolute value.
In all other cases (i.e.\ if the~$\lambda^{xy}_j$ are not all real and do not all 
have the same absolute value),
the points~$x$ and~$y$ are said to be {\em{lightlike}} separated. }}
\end{Def} \noindent
Restricting the causal structure of~$\F$ to~$M$, we get causal relations in spacetime.

The Lagrangian~\eqref{Lagrange} is compatible with the above notion of causality in the
following sense.
Suppose that two points~$x, y \in M$ are spacelike separated.
Then the eigenvalues~$\lambda^{xy}_i$ all have the same absolute value.
As a consequence, the Lagrangian~\eqref{Lagrange} vanishes. Thus pairs of points with spacelike
separation do not enter the action. This can be seen in analogy to the usual notion of causality where
points with spacelike separation cannot influence each other.
This is the reason for the notion ``causal'' in {\em{causal}} fermion system
and {\em{causal}} action principle.

Moreover, a causal fermion system distinguishes a {\em{direction of time}}.
To this end, we let~$\pi_x$ be the orthogonal projection in~$\H$ on the subspace~$x(\H) \subset \H$
and introduce the functional
\[ 
{\mathscr{C}} \::\: M \times M \rightarrow \R\:,\qquad
{\mathscr{C}}(x, y) := i \tr \big( y\,x \,\pi_y\, \pi_x - x\,y\,\pi_x \,\pi_y \big) \:. \]
Obviously, this functional is anti-symmetric in its two arguments, making it possible to introduce the notions
\[ 
\left\{ \begin{array}{cl} \text{$y$ lies in the {\em{future}} of~$x$} &\quad \text{if~${\mathscr{C}}(x, y)>0$} \\[0.2em]
\text{$y$ lies in the {\em{past}} of~$x$} &\quad \text{if~${\mathscr{C}}(x, y)<0$}\:. \end{array} \right. \]

\subsection{The Euler-Lagrange Equations} \label{secEL}
A minimizer of a causal variational principle
satisfies the following {\em{Euler-Lagrange (EL) equations}}.
For a suitable value of the parameter~$\s>0$,
the function~$\ell : \F \rightarrow \R_0^+$ defined by
\beq \label{elldef}
\ell(x) := \int_M \L_\kappa(x,y)\: d\rho(y) - \s
\eeq
is minimal and vanishes on spacetime~$M:= \supp \rho$,
\beq \label{EL}
\ell|_M \equiv \inf_\F \ell = 0 \:.
\eeq
Here the $\kappa$-{\em{Lagrangian}}~$\L_\kappa$ is defined by
\beq \label{Lkappa}
\L_\kappa \::\: \F \times \F \rightarrow \R\:,\qquad
\L_\kappa(x,y) := \L(x,y) + \kappa\: |xy|^2
\eeq
with a non-negative parameter~$\kappa$, which can be thought of as the Lagrange parameter
corresponding to the boundedness constraint.
Likewise, the parameter~$\s \geq 0$ in~\eqref{elldef} is the Lagrange parameter
corresponding to the volume constraint. For the derivation and further details we refer to~\cite[Section~2]{jet}
or~\cite[Chapter~7]{intro}.

\subsection{Spinors and Physical Wave Functions} \label{secinherent}
A causal fermion system also gives rise to spinorial wave functions in spacetime, as we now explain.
For every~$x \in \F$ we define the {\em{spin space}}~$S_x$ by~$S_x = x(\H)$;
it is a subspace of~$\H$ of dimension at most~$2n$.
It is endowed with the {\em{spin inner product}} $\Sl .|. \Sr_x$ defined by
\beq \label{ssp}
\Sl u | v \Sr_x = -\la u | x v \ra_\H \qquad \text{(for all $u,v \in S_x$)} \:.
\eeq
A {\em{wave function}}~$\psi$ is defined as a function
which to every~$x \in M$ associates a vector of the corresponding spin space,
\[ 
\psi \::\: M \rightarrow \H \qquad \text{with} \qquad \psi(x) \in S_xM \quad \text{for all~$x \in M$}\:. \]

It is an important observation that every vector~$u \in \H$ of the Hilbert space gives rise to a unique
wave function denoted by~$\psi^u$. It is obtained by projecting the vector~$u$
to the corresponding spin spaces,
\beq \label{psiudef}
\psi^u \::\: M \rightarrow \H\:,\qquad \psi^u(x) := \pi_x u \in S_xM \:.
\eeq
We refer to~$\psi^u$ as the {\em{physical wave function}} of the vector~$u \in \H$.
Choosing an orthonormal basis~$(e_i)$ of~$\H$, we obtain a whole family of
physical wave functions~$(\psi^{e_i})$. This ensemble of wave functions is crucial
for the understanding of what a causal fermion systems is about.
In fact, all spacetime structures (like for example the causal structure in Definition~\ref{def2})
can be recovered from this ensemble. Moreover, one can construct concrete examples
of causal fermion systems by choosing the physical wave functions more specifically as
the quantum mechanical wave functions
in a classical Lorentzian spacetime. In the next section we explain this construction in more detail.

\section{The Limiting Case of Classical Gravity} \label{secclass}
In this section we outline how and in which sense the causal action principle
gives rise to classical gravity. For more details we refer to~\cite{cfs} and the
review article~\cite{nrstg}.

\subsection{Describing a Lorentzian Spacetime by a Causal Fermion System} \label{seccfslorentz}
We now explain how a classical curved spacetime is described by a causal fermion system.
Our starting point is Lorentzian spin geometry.
Thus we let~$(\scrM, g)$ be a smooth, globally hyperbolic, time-oriented
Lorentzian spin manifold of dimension four.
For the signature of the metric we use the convention~$(+ ,-, -, -)$.
We denote the corresponding spinor bundle by~$S\scrM$. Its fibers~$S_x\scrM$ (with~$x \in \scrM$)
are endowed with an inner product~$\Sl .|. \Sr_x$ of signature~$(2,2)$.
Clifford multiplication is described by a mapping~$\gamma$
which satisfies the anti-commutation relations,
\[ 
\gamma \::\: T_x\scrM \rightarrow \Lin(S_x\scrM) \qquad
\text{with} \qquad \gamma(u) \,\gamma(v) + \gamma(v) \,\gamma(u) = 2 \, g(u,v)\,\1_{S_x(\scrM)} \:. \]
We also write Clifford multiplication in components with the Dirac matrices~$\gamma^j$.
The metric connections on the tangent bundle and the spinor bundle are denoted by~$\nabla$.
The sections of the spinor bundle are also referred to as wave functions.

We denote the smooth sections of the spinor bundle by~$C^\infty(\scrM, S\scrM)$.
The Dirac operator~$\Dir$ is defined by
\[ \Dir := i \gamma^j \nabla_j \::\: C^\infty(\scrM, S\scrM) \rightarrow C^\infty(\scrM, S\scrM)\:. \]
Given a real parameter~$m \in \R$ (the {\em{mass}}), the Dirac equation reads
\[ (\Dir - m) \,\psi = 0 \:. \]
We mainly consider solutions in the class~$\Cisc(\scrM, S\scrM)$ of smooth sections
with spatially compact support (i.e.\ wave functions whose restriction to any Cauchy surface
is compact). On such solutions, one has the scalar product
\[ 
(\psi | \phi)_m = 2 \pi \int_\scrN \Sl \psi \,|\, \gamma(\nu)\, \phi \Sr_x\: d\mu_\scrN(x) \:, \]
where~$\scrN$ denotes any Cauchy surface and~$\nu$ its future-directed normal
(due to current conservation, the scalar product is
in fact independent of the choice of~$\scrN$; for details see~\cite[Section~2]{finite}).
Forming the completion gives the Hilbert space~$(\H_m, (.|.)_m)$.

Next, we choose a closed subspace~$\H \subset \H_m$
of the solution space of the Dirac equation.
The induced scalar product on~$\H$ is denoted by~$\la .|. \ra_\H$.
There is the technical difficulty that the wave functions in~$\H$ are in general not continuous,
making it impossible to evaluate them pointwise.
For this reason, we need to introduce an {\em{ultraviolet regularization}} on
the length scale~$\varepsilon$, described mathematically by a linear
\[ \text{\em{regularization operator}} \qquad {\mathfrak{R}}_\varepsilon \::\: \H \rightarrow C^0(\scrM, S\scrM) \:. \]
In the simplest case, the regularization can be realized by a convolution
on a Cauchy surface or in spacetime (for details see~\cite[Section~4]{finite}
or~\cite[Section~\S1.1.2]{cfs}). For us, the regularization is not merely a technical tool,
but it realizes the concept that we want to change the geometric structures on the microscopic
scale. With this in mind, we always consider the regularized quantities as those having mathematical and
physical significance. Different choices of regularization operators realize different
microscopic spacetime structures.

Evaluating the regularization operator at a spacetime point~$x \in \scrM$ gives 
the {\em{regularized wave evaluation operator}}~$\Psi^\varepsilon(x)$,
\beq \label{Pepsdirac}
\Psi^\varepsilon(x) = {\mathfrak{R}}_\varepsilon(x) \::\: \H \rightarrow S_x \scrM \:.
\eeq
We also take its adjoint (with respect to the Hilbert space
scalar product~$\la .|. \ra_\H$ and the spin inner product~$\Sl .|. \Sr_x$),
\[ \big(\Psi^\varepsilon(x) \big)^* \::\: S_x \scrM \rightarrow \H  \:. \]
Multiplying~$\Psi^\varepsilon(x)$ by its adjoint gives the operator
\beq \label{Fepsprod}
F^\varepsilon(x) := - \big( \Psi^\varepsilon(x) \big)^* \,\Psi^\varepsilon(x) \::\: \H \rightarrow \H \:,
\eeq
referred to as the {\em{local correlation operator}} at
the spacetime point~$x$. The local correlation operator is also characterized by the relation
\beq \label{Fepsdef}
(\psi \,|\, F^\varepsilon(x)\, \phi) = -\Sl ({\mathfrak{R}}_\varepsilon\psi)(x) | 
({\mathfrak{R}}_\varepsilon \phi)(x) \Sr_x \qquad \text{for all~$\psi, \phi
\in \H$} \:.
\eeq
Taking into account that the inner product on the Dirac spinors at~$x$ has signature~$(2,2)$,
it is a symmetric operator on~$\H$
of rank at most four, which (counting multiplicities) has at most two positive and at most two negative eigenvalues.
Varying the spacetime point, we obtain a mapping
\[ F^\varepsilon \::\: \scrM \rightarrow \F \subset \Lin(\H)\:, \]
where~$\F$ denotes all symmetric operators of
rank at most four with at most two positive and at most two negative eigenvalues.
Finally, we introduce the measure~$\rho$ on~$\F$ by taking
the push-forward of the volume measure on~$\scrM$ under the mapping~$F^\varepsilon$,
\beq \label{rhoeps}
\rho := (F^\varepsilon)_* \mu_\scrM
\eeq
(thus~$\rho(\Omega) := \mu_\scrM((F^\varepsilon)^{-1}(\Omega))$).
The resulting structure~$(\H, \F, \rho)$ is a causal fermion system of spin dimension two.

We conclude with a few comments on the significance of this construction.
We first point out that the construction uses all the structures of Lorentzian spin geometry
as well as the properties of the Dirac wave functions in~$\H$.
We thus obtain a {\em{very specific class of examples}} of causal fermion systems
describing classical spacetimes. In general, the measure~$\rho$ defined by~\eqref{rhoeps}
will not be a minimizer of the causal action principle. But it is an approximate minimizer
if the Einstein equations are satisfied. Before making this point precise in the next section
(Section~\ref{seccl}), we now explain the physical picture behind the Dirac wave functions in~$\H$.
These wave functions have the interpretation as 
being those Dirac wave functions which are realized in the physical system under
consideration. If we describe for example a system of one electron,
then the wave function of the electron is contained in~$\H$.
Moreover, $\H$ includes all the wave functions
which form the so-called Dirac sea (for an explanation of this point
see for example~\cite{srev}). All the Dirac wave functions in~$\H$
can be identified with the ensemble of {\em{physical wave functions}} of the causal fermion system
as introduced in~\eqref{psiudef}. This ``identification'' is made mathematically precise by
also identifying the objects of Lorentzian spin geometry with corresponding inherent objects of the causal
fermion system (for details see~\cite[Section~1.2]{intro}).
The name causal {\em{fermion}} system is motivated by the fact that Dirac particles are fermions.
According to~\eqref{Fepsdef}, the local correlation operator~$F^\varepsilon(p)$ describes 
densities and correlations of the physical
wave functions at the spacetime point~$p$.
Working exclusively with the local correlation operators and the
corresponding push-forward measure~$\rho$ means in particular
that the geometric structures are encoded in and must be retrieved from the physical wave functions.
Since the physical wave functions describe the distribution of
matter in spacetime, one can summarize this concept
by saying that {\em{matter encodes geometry}}.

\subsection{Classical Gravity in the Continuum Limit} \label{seccl}
The construction of the causal fermion system in the previous section involved the
Lorentzian metric~$g$. But this Lorentzian metric did not need to satisfy the
Einstein equation. Instead of postulating the Einstein equations, our strategy is to {\em{derive}}
these equations from the causal action principle. To this end, we need to evaluate the
EL equations~\eqref{elldef} for the causal fermion system~$(\H, \F, \rho)$ constructed in the
previous section in the limit~$\varepsilon \searrow 0$ when the ultraviolet
regularization is removed. This analysis, referred to as the {\em{continuum limit}}, is carried out
in detail in~\cite[Chapter~4]{cfs}. Before giving a brief outline of how this analysis works,
we state the main result in the context of gravitational fields: The EL equations~\eqref{elldef} are satisfied
asymptotically for small~$\varepsilon>0$ only if the Lorentzian metric satisfies the Einstein equations,
up to possible higher order corrections in curvature (which scale in powers of~$(\delta^2\:
\text{Riem})$, where~$\delta$ is the Planck length and~$\text{Riem}$ is the
curvature tensor), i.e.\ (see~\cite[Theorems~4.9.3 and~5.4.4]{cfs})
\beq \label{einstein}
R_{jk} - \frac{1}{2}\:R\: g_{jk} + \Lambda\, g_{jk} = G\, T_{jk} 
+ {\mathscr{O}} \big( \delta^4 \,\text{Riem}^2 \big) \:.
\eeq
Moreover, it is shown that the gravitational coupling constant~$G$ is determined by
the length scale of the microscopic spacetime structures and has the scaling
\[ G \sim \delta^2 \:. \]
The cosmological constant~$\Lambda$, however, is not determined by our method.
In order to avoid confusion, we finally note that we carefully distinguish the Planck length~$\delta$
from the regularization length~$\varepsilon$. The reason is that, although it seems natural to
assume that these length scales coincide, this does not necessarily need to be the case.
In fact, the are indications that~$\varepsilon$ should be chosen even much smaller than~$\delta$
(for a detailed discussion of the length scales see for example~\cite[Appendix~A]{jacobson}).

We now briefly explain the general procedure in the analysis of the continuum limit.
Given a causal fermion system~$(\H, \F, \rho)$ describing a globally hyperbolic spacetime~$(\scrM, g)$
(as constructed in Section~\ref{seccfslorentz}), our task is to evaluate the EL equations~\eqref{EL}.
The first step is to compute the eigenvalues~$\lambda^{xy}_1, \ldots, \lambda^{xy}_{2n}$ of the operator product~$F^\varepsilon(x)\, F^\varepsilon(y)$
for given~$x,y \in \scrM$. Using~\eqref{Fepsprod} together with the fact that the non-zero eigenvalues
of a matrix product as well as the corresponding algebraic multiplicities
do not change when the matrices are cyclically commuted,
one can just as well compute the eigenvalues of the {\em{closed chain}}~$A_{xy}$ defined by
\beq \label{closedchain}
A_{xy} := P^\varepsilon(x,y) \, P^\varepsilon(y,x) \::\: S_x\scrM \rightarrow S_x\scrM \:,
\eeq
where the {\em{regularized kernel of the fermionic projector}}~$P^\varepsilon(x,y)$ is defined by
\[ P^\varepsilon(x,y) = -\Psi^\varepsilon(x)\big( \Psi^\varepsilon(y) \big)^* 
 \::\: S_y\scrM \rightarrow S_x\scrM \:. \]
In this way, it suffices to compute the eigenvalues of a linear endomorphism of~$S_xM$,
which can be represented by a~$4 \times 4$-matrix. Moreover, the regularized kernel of the fermionic projector
can be analyzed explicitly using the {\em{regularized Hadamard expansion}}.
In general terms, $P^\varepsilon(x,y)$ is a smooth function which in the limit~$\varepsilon \searrow 0$
converges to a distribution which has singularities on the lightcone
(i.e.\ for lightlike separation of~$x$ and~$y$).
As a consequence, the pointwise product in~\eqref{closedchain} is ill-defined in this limit.
In the continuum limit analysis, one deals with this issue by studying the Lagrangian and other
composite expressions in~$P^\varepsilon(x,y)$ asymptotically for small~$\varepsilon>0$.
Evaluating the resulting expressions in the EL equations~\eqref{EL},
one gets equations involving the gravitational field and the Dirac wave functions.
In models containing neutrinos, these equations imply the Einstein equations~\eqref{einstein}.

The detailed computations can be found in~\cite{cfs}, where the continuum limit analysis is
carried out in Minkowski space. In this context, the Hadamard expansion
is referred to as the {\em{light cone expansion}}.
The reader interested in the general construction of the regularized Hadamard expansion
in curved spacetime is referred to~\cite{reghadamard}.

\section{Going Beyond Classical Gravity} \label{secbeyond}
The above derivation of the Einstein equations in the continuum limit has two disadvantages.
First, it is rather technical and thus does not give a good intuitive
understanding of the underlying mechanisms.
Second and more importantly, the Einstein equations are obtained
only for the special class of examples of causal fermion systems constructed in Section~\ref{seccfslorentz}
and hold only in the continuum limit. But the methods do not give an insight into the geometric meaning
of the EL equations~\eqref{EL} for general causal fermion systems describing
more general ``quantum'' spacetimes. This raises the following question:
\beq
\begin{split}
&\text{Given a general causal fermion system~$(\H, \F, \rho)$, how do the} \\[-0.4em]
&\text{EL equations~\eqref{EL} relate matter to the geometry of spacetime?}
\end{split} \label{question}
\eeq
For a general causal fermion system, we cannot work with tensor fields, making it impossible
to formulate the Einstein equations or modifications thereof.
Therefore, we need to go beyond the mathematical setting of Lorentzian geometry.
In the next sections, we explain step by step how this can be done and mention
a few results.

\subsection{A Lorentzian Quantum Geometry} \label{seclqg}
In~\cite{lqg} it was shown that a causal fermion system gives rise to geometric structures
in spacetime. The general strategy for obtaining these geometric structures is as follows.
Given two spacetime points~$x,y \in M$, the corresponding spin space~$S_x$ and~$S_y$
(see Section~\ref{secinherent}) are subspaces of the underlying Hilbert space~$\H$.
Denoting the orthogonal projection to the spin spaces~$S_x$ again by
\[ \pi_x : \H \rightarrow S_x \:, \]
the {\em{kernel of the fermionic projector}} is introduced by
\[ P(x,y) = \pi_x y|_{S_y} \::\: S_y \rightarrow S_x \:. \]
Being an operator from one spin space to another, it gives relations between the spacetime points.
In other words, the kernel of the fermionic projector induces additional structures in spacetime.
One important structure is the {\em{spin connection}}~$D_{x,y}$, being a unitary mapping between
the spin spaces,
\beq \label{Dxy}
D_{x,y} \::\: S_yM \rightarrow S_xM \qquad \text{unitary}
\eeq
(unitary with respect to the spin inner product~\eqref{ssp}).
A first idea for construction~$D_{x,y}$ is to take a polar decomposition of~$P(x,y)$.
This idea needs to be refined in order to also obtain a metric connection and to arrange
that the different connections are compatible. Here for brevity we omit the details and refer
to~\cite{lqg} or the review~\cite{nrstg}. In general terms, it turns out that there is a canonical
spin connection~\eqref{Dxy}, provided that the operators~$x$ and~$y$ satisfy certain conditions,
which are subsumed in the notion that the spacetime points be {\em{spin-connectable}}.
{\em{Curvature}}~$\mathfrak{R}$ can be defined as the holonomy of the spin connection.
Thus, in the simplest case, for three points~$x,y,z \in M$ which are mutually spin-connectable,
one sets
\[ \mathfrak{R}(x,y,z) = D_{x,y}\: D_{y,z}\:D_{z,x} \::\: S_xM \rightarrow S_xM \:. \]
In~\cite{lqg} also the correspondence to Lorentzian spin geometry is established,
in the sense that for causal fermion systems describing a globally hyperbolic
Lorentzian manifold (as constructed in Section~\ref{seccfslorentz})
and taking a suitable limit~$\varepsilon \searrow 0$,
the spin connection~\eqref{Dxy} goes over to the spinorial Levi-Civita connection of the Lorentzian manifold.

While these structures give a good understanding of the geometry of a causal fermion system,
so far they have not been fruitful for unraveling the form of the gravitational interaction
as described by the causal action principle. The basic problem is that the EL equations~\eqref{EL}
cannot be formulated in terms of these geometric structures.
In other words, it does not seem possible to rewrite the causal action principle as
a geometric variational principle involving~$D_{x,y}$ and~$\mathfrak{R}$.
With this in mind, we now move on to other structures which again have a geometric meaning,
but harmonize better with the EL equations~\eqref{EL}.

\subsection{Surface Layer Integrals} \label{secsli}
Coming back to our question~\eqref{question}, we follow another path
for getting a connection between the geometric structures and the EL equations~\eqref{EL}.
This method is inspired by the fact that the effects of gravity can also be captured by
considering the volume and area of surfaces in spacetime, and by analyzing how this area
changes under flows of the surfaces. Typical examples for this connection are Huisken's
isoperimetric mass (see for example~\cite{jauregui-lee}) and Jacobson's connection between area
change and matter flux~\cite{jacobson}.
Thinking along these lines, the first obvious question is how a surface integral can be
defined in the setting of causal fermion systems. Once this question has been answered,
one can analyze the area of families of surfaces and make the above analogies more precise.

As a typical example, suppose we want to define the analog of an integral over a Cauchy surface~$\scrN$
in a globally hyperbolic spacetime, i.e.\ symbolically
\beq \label{conserve}
\int_\scrN (\cdots)\: d\mu_{\scrN}(x) \:,
\eeq
where~$d\mu_\scrN$ is the induced volume measure on~$\scrN$. Here one can
think of~$(\cdots)$ as a density like for example the inner product~$\nu_i J^i$ of a current vector
field with the future-directed normal~$\nu$.
In the setting of causal fermion system, surface integrals like~\eqref{conserve}
are undefined. Instead, one considers so-called {\em{surface layer integrals}}.
In general terms, a surface layer integral is a double integral of the form
\beq \label{intdouble}
\int_\Omega \bigg( \int_{M \setminus \Omega} (\cdots)\: \L_\kappa(x,y)\: d\rho(y) \bigg)\, d\rho(x) \:,
\eeq
where one variable is integrated over a subset~$\Omega \subset M$, and the other
variable is integrated over the complement of~$\Omega$. Here~$(\cdots)$ is the analog of
the corresponding factor in~\eqref{conserve}, but now having the mathematical structure of
being a differential operator acting on the Lagrangian.

\begin{figure}
\psscalebox{1.0 1.0} 
{
\begin{pspicture}(0,-1.511712)(10.629875,1.511712)
\definecolor{colour0}{rgb}{0.8,0.8,0.8}
\definecolor{colour1}{rgb}{0.6,0.6,0.6}
\pspolygon[linecolor=black, linewidth=0.002, fillstyle=solid,fillcolor=colour0](6.4146066,0.82162136)(6.739051,0.7238436)(6.98794,0.68384355)(7.312384,0.66162133)(7.54794,0.67939913)(7.912384,0.7593991)(8.299051,0.8705102)(8.676828,0.94162136)(9.010162,0.9549547)(9.312385,0.9371769)(9.690162,0.8571769)(10.036829,0.7371769)(10.365718,0.608288)(10.614607,0.42162135)(10.614607,-0.37837866)(6.4146066,-0.37837866)
\pspolygon[linecolor=black, linewidth=0.002, fillstyle=solid,fillcolor=colour1](6.4146066,1.2216214)(6.579051,1.1616213)(6.770162,1.1127324)(6.921273,1.0905102)(7.103495,1.0816213)(7.339051,1.0549546)(7.530162,1.0638436)(7.721273,1.0993991)(7.8857174,1.1393992)(8.10794,1.2060658)(8.299051,1.2549547)(8.512384,1.3038436)(8.694607,1.3260658)(8.890162,1.3305103)(9.081273,1.3393991)(9.379051,1.3216213)(9.659051,1.2593992)(9.9746065,1.1705103)(10.26794,1.0460658)(10.459051,0.94384354)(10.614607,0.82162136)(10.610162,0.028288014)(10.414606,0.1660658)(10.22794,0.26828802)(10.010162,0.37051025)(9.663495,0.47273245)(9.356829,0.53051025)(9.054606,0.548288)(8.814607,0.54384357)(8.58794,0.5171769)(8.387939,0.48162135)(8.22794,0.44162133)(7.90794,0.34828803)(7.6946063,0.29939914)(7.485718,0.26828802)(7.272384,0.26828802)(7.02794,0.28162134)(6.82794,0.3171769)(6.676829,0.35273245)(6.543495,0.38828802)(6.4146066,0.42162135)
\pspolygon[linecolor=black, linewidth=0.002, fillstyle=solid,fillcolor=colour0](0.014606438,0.82162136)(0.3390509,0.7238436)(0.5879398,0.68384355)(0.9123842,0.66162133)(1.1479398,0.67939913)(1.5123842,0.7593991)(1.8990508,0.8705102)(2.2768288,0.94162136)(2.610162,0.9549547)(2.9123843,0.9371769)(3.290162,0.8571769)(3.6368287,0.7371769)(3.9657176,0.608288)(4.2146063,0.42162135)(4.2146063,-0.37837866)(0.014606438,-0.37837866)
\psbezier[linecolor=black, linewidth=0.04](6.4057174,0.8260658)(7.6346064,0.45939913)(7.8634953,0.8349547)(8.636828,0.92828804)(9.410162,1.0216213)(10.165717,0.7927325)(10.614607,0.42162135)
\psbezier[linecolor=black, linewidth=0.04](0.005717549,0.8260658)(1.2346064,0.45939913)(1.4634954,0.8349547)(2.2368286,0.92828804)(3.0101619,1.0216213)(3.7657175,0.7927325)(4.2146063,0.42162135)
\rput[bl](2.0101619,0.050510235){$\Omega$}
\rput[bl](8.759051,0.0016213481){\normalsize{$\Omega$}}
\psline[linecolor=black, linewidth=0.04, arrowsize=0.09300000000000001cm 1.0,arrowlength=1.7,arrowinset=0.3]{->}(1.9434953,0.85495466)(1.8057176,1.6193991)
\rput[bl](2.0946064,1.1705103){$\nu$}
\psbezier[linecolor=black, linewidth=0.02](6.4146066,0.42384356)(7.6434956,0.057176903)(7.872384,0.43273246)(8.645718,0.52606577)(9.419051,0.61939913)(10.174606,0.39051023)(10.623495,0.019399125)
\psbezier[linecolor=black, linewidth=0.02](6.410162,1.2193991)(7.639051,0.8527325)(7.86794,1.228288)(8.6412735,1.3216213)(9.414606,1.4149547)(10.170162,1.1860658)(10.619051,0.8149547)
\rput[bl](8.499051,0.9993991){\normalsize{$y$}}
\rput[bl](7.8657174,0.49273247){\normalsize{$x$}}
\psdots[linecolor=black, dotsize=0.06](8.170162,0.65273243)
\psdots[linecolor=black, dotsize=0.06](8.796828,1.1327325)
\psline[linecolor=black, linewidth=0.02](6.1146064,1.2216214)(6.103495,0.82162136)
\rput[bl](5.736829,0.8993991){\normalsize{$l$}}
\rput[bl](3.6146064,0.888288){$\scrN$}
\rput[bl](1.1146064,-1.4117119){$\displaystyle \int_\scrN \cdots\, d\mu_\scrN$}
\rput[bl](5.7146063,-1.511712){$\displaystyle \int_\Omega d\rho(x) \int_{M \setminus \Omega} d\rho(y)\: \cdots\:\L_\kappa(x,y)$}
\psline[linecolor=black, linewidth=0.02](6.0146065,1.2216214)(6.2146063,1.2216214)
\psline[linecolor=black, linewidth=0.02](6.0146065,0.82162136)(6.2146063,0.82162136)
\end{pspicture}
}
\caption{A surface integral and a corresponding surface layer integral.}
\label{fignoether1}
\end{figure}
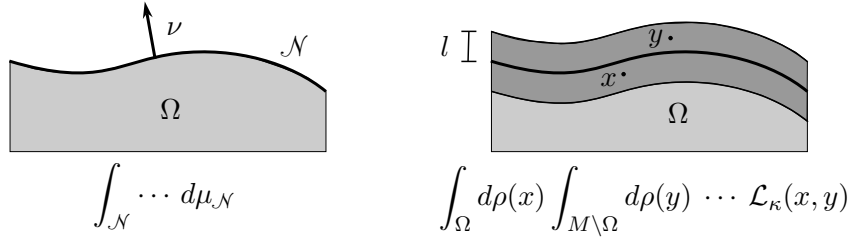

In order to explain the basic concept, let us assume for a moment that
the Lagrangian is of {\em{short range}}
in the following sense.
We let~$d \in C^0(M \times M, \R^+_0)$ be a suitably chosen distance function on~$M$. Then the
assumption of short range can be quantified by demanding that~$\L_\kappa$ should vanish
on distances larger than~$l$, i.e.
\beq \label{shortrange}
d(x,y) > l \quad \Longrightarrow \quad \L_\kappa(x,y) = 0 \:.
\eeq
Under this assumption,
the surface layer integral~\eqref{intdouble} only involves pairs~$(x,y)$ of distance at most~$l$,
where~$x$ lies in~$\Omega$, whereas~$y$ lies in the complement~$M \setminus \Omega$.
As a consequence, the integral only involves points in a layer around the boundary of~$\Omega$
of width~$l$, i.e.
\[ x, y \in B_l \big(\partial \Omega \big) \:. \]
Therefore, a double integral of the form~\eqref{intdouble} can be regarded as an approximation
of a surface integral on the length scale~$l$, as shown in Figure~\ref{fignoether1}.
In most applications, the Lagrangian is {\em{not}} of short range in the strict sense~\eqref{shortrange}.
But it decays on the Compton scale~$l \sim 1/m$ (where~$m$ denotes again the
mass of the Dirac particles), so that the qualitative picture in Figure~\ref{fignoether1} still applies.

\begin{figure}[tb]
\psscalebox{1.0 1.0} 
{
\begin{pspicture}(0,26.0)(9.62,29.2)
\definecolor{colour3}{rgb}{0.7019608,0.7019608,0.7019608}
\definecolor{colour4}{rgb}{0.9019608,0.9019608,0.9019608}
\pspolygon[linecolor=colour3, linewidth=0.02, fillstyle=solid,fillcolor=colour3](6.0300636,27.62005)(6.2800636,27.64005)(6.7000637,27.72005)(7.1200633,27.80005)(7.6400633,27.83005)(8.080064,27.810051)(8.570064,27.67005)(8.960064,27.61005)(9.295063,27.60005)(9.420063,27.57005)(9.610064,27.62005)(9.620064,26.04005)(6.0200634,26.01005)
\pspolygon[linecolor=colour4, linewidth=0.02, fillstyle=solid,fillcolor=colour4](6.0300636,29.23005)(9.610064,29.22505)(9.610064,27.64005)(9.395063,27.60005)(9.090063,27.58505)(8.760063,27.62005)(8.330064,27.78005)(7.7700634,27.86005)(7.7700634,27.86005)(7.5400634,27.85005)(7.1000633,27.82005)(6.7700634,27.76005)(6.4100633,27.68005)(6.2000637,27.66005)(6.0300636,27.66005)
\pspolygon[linecolor=colour3, linewidth=0.02, fillstyle=solid,fillcolor=colour3](0.020063477,26.02005)(3.6200635,26.02005)(3.6150634,27.59505)(3.4700634,27.55005)(3.1900635,27.51005)(2.9100635,27.53005)(2.5700636,27.57005)(2.2100635,27.66005)(1.7900635,27.73005)(1.4600635,27.76005)(1.1200634,27.79005)(0.84006345,27.77005)(0.4600635,27.73005)(0.20006348,27.68005)(0.030063476,27.61005)
\pspolygon[linecolor=colour4, linewidth=0.02, fillstyle=solid,fillcolor=colour4](0.020063477,29.22005)(3.6100636,29.22005)(3.6150634,27.63005)(3.4400635,27.560051)(3.1500635,27.53005)(2.6900635,27.58005)(2.1300635,27.70005)(1.6800635,27.77005)(1.2000635,27.83005)(0.64006346,27.78005)(0.24006347,27.72005)(0.030063476,27.64005)
\psline[linecolor=black, linewidth=0.04, arrowsize=0.05291667cm 2.0,arrowlength=1.4,arrowinset=0.0]{<->}(3.4200635,28.63005)(6.4200635,26.63005)
\psline[linecolor=black, linewidth=0.04, arrowsize=0.05291667cm 2.0,arrowlength=1.4,arrowinset=0.0]{<->}(2.8200636,26.83005)(6.8200636,28.43005)
\psbezier[linecolor=black, linewidth=0.04](0.020063477,27.63005)(0.24502629,27.729553)(0.6800635,27.84005)(1.4800634,27.780050048828127)(2.2800634,27.72005)(3.0623415,27.364084)(3.6200635,27.63005)
\psbezier[linecolor=black, linewidth=0.04](6.0200634,27.63005)(6.8838763,27.696981)(6.8200636,27.84005)(7.7300634,27.840050048828125)(8.640063,27.84005)(8.806251,27.39698)(9.620064,27.63005)
\rput[bl](7.3,28){$M \setminus \Omega$}
\rput[bl](7.7,26.8){$\Omega$}
\rput[bl](1.3,28){$\tilde{M} \setminus \tilde{\Omega}$}
\rput[bl](1.7,26.8){$\tilde{\Omega}$}
\rput[bl](2.55,26.7){$x$}
\rput[bl](3.15,28.5){$y$}
\rput[bl](6.5,26.5){$x$}
\rput[bl](6.9,28.3){$y$}
\end{pspicture}
}
\caption{The nonlinear surface layer integral.}
\label{figosinl}
\end{figure}
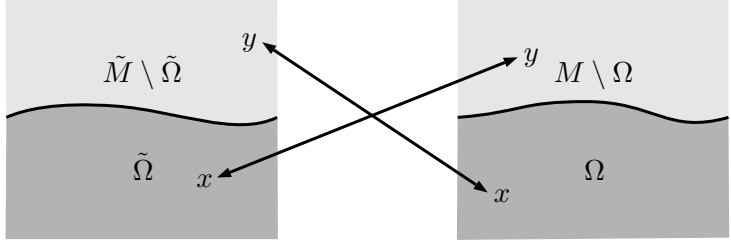

Surface layer integrals were first introduce in~\cite{noether} in the context of Noether-like theorems.
The analysis of this paper revealed that, in contrast to the geometric structures in Section~\ref{seclqg},
the structure of a surface layer integral does fit nicely to the EL equations~\eqref{EL},
giving rise to very useful conservation laws. This connection was analyzed further and more systematically
in~\cite{jet, osi}. Moreover, in~\cite{fockbosonic} another variant of a surface layer integral was
introduced, which will be importance for what follows.
This new surface layer integral can be regarded as a generalization or ``nonlinear version'' of~\eqref{intdouble},
as can be understood as follows. The differential operator~$(\cdots)$ in
the integrand can be regarded as describing
first or second variations of the measure~$\rho$.
Instead of considering variations of~$\rho$, 
we now consider an additional measure~$\tilde{\rho}$ which can be thought of as a finite
perturbation of the measure~$\rho$. Consequently, we also have two spacetimes
\[ M := \supp \rho \qquad \text{and} \qquad \tilde{M} := \supp \tilde{\rho} \:. \]
Choosing two compact subsets~$\Omega \subset M$ and~$\tilde{\Omega} \subset \tilde{M}$
of the corresponding spacetimes, we form the {\em{nonlinear surface layer integral}} by
\beq \label{osinl}
\gamma^{\tilde{\Omega}, \Omega}(\tilde{\rho}, \rho) :=
\int_{\tilde{\Omega}} d\tilde{\rho}(x) \int_{M \setminus \Omega} d\rho(y)\: \L_\kappa(x,y)
- \int_{\Omega} d\rho(x) \int_{\tilde{M} \setminus \tilde{\Omega}}  d\tilde{\rho}(y)\: \L_\kappa(x,y) \:.
\eeq
Note that one argument of the Lagrangian is in~$M$, whereas the other is in~$\tilde{M}$.
Moreover, one argument lies inside the set~$\Omega$ respectively $\tilde{\Omega}$,
whereas the other argument lies outside this set.
In this way, the nonlinear surface layer integral ``compares'' the two spacetimes
near the boundaries of~$\Omega$ and~$\tilde{\Omega}$,
as is illustrated in Figure~\ref{figosinl}.

\subsection{The Total Mass of a Static Causal Fermion System} \label{secpmt}
Combining the general structure of Huisken's isoperimetric mass with the
concept of a surface layer integral, the question arises whether notions like the total mass and total energy
can be introduced for a class of causal fermion systems which generalize asymptotically flat Lorentzian
manifolds. This question has been answered in the affirmative in~\cite{pmt}
in the static setting. We now outline a few constructions and results of this paper.

Static causal fermion systems are introduced as usual by demanding a one-parameter 
group of symmetries:
\begin{Def} \label{defstatic}
Let~$(\scrU_t)_{t \in \R}$ be a strongly continuous one-parameter group of unitary transformations 
on the Hilbert space~$\H$ (i.e.\ $s$-$\lim_{t' \rightarrow t} \scrU_{t'}= \scrU_t$ and~$\scrU_t \scrU_{t'} = \scrU_{t+t'}$).
The causal fermion system~$(\H, \F, \rho)$ is {\bf{static with respect to~$(\scrU_t)_{t \in \R}$}}
if it has the following properties:
\begin{itemize}[leftmargin=2em]
\item[\rm{(i)}] Spacetime $M:= \supp \rho \subset \F$ is a topological product,
\[ M = \R \times N \:. \]
We write a spacetime point~$x \in M$ as~$x=(t,\x)$ with~$t \in \R$ and~$\x \in N$.
\item[\rm{(ii)}] The one-parameter group~$(\scrU_t)_{t \in \R}$ leaves
the measure~$\rho$ invariant, i.e.
\[ \rho\big( \scrU_t \,\Omega\, \scrU_t^{-1} \big) = \rho(\Omega) \qquad \text{for all
	$\rho$-measurable~$\Omega \subset \F$} \:. \]
Moreover,
\[ \scrU_{t'}\: (t,\x)\: \scrU_{t'}^{-1} = (t+t',\x)\:. \]
\end{itemize}
\end{Def} \noindent
As a consequence of~(ii), the measure~$\rho$ has the product form
\[ 
d\rho = dt \,d\mu \:, \]
where~$\mu$ is a measure on~$\G:=\G = \F^\text{reg} / \R$.

Restricting attention to static causal fermion systems and varying within the class
of measures which are invariant under the action of a given one-parameter group~$(\scrU_t)_{t \in \R}$,
one obtains a corresponding {\em{static causal action principle}}, where one minimizes the static action
\[ 
\Sact_\text{static} (\mu) = \int_\G d\mu(\x) \int_\G d\mu(\y)\: \L_\text{static}(\x,\y) \]
under variations of the measure~$\mu$ on~$\G$, keeping the total volume~$\mu(\G)$ fixed
({\em{volume constraint}}). The static Lagrangian is obtained by integrating
one argument of the $\kappa$-Lagrangian~\eqref{Lkappa} over time,
\[ \L_\text{static}(\x,\y) := \int_{-\infty}^\infty \L_\kappa\big( (0,\x), (t, \y) \big) \: dt \:. \]

The {\em{total mass}}~$\Mass$ of a static causal fermion system described by the measure~$\tilde{\mu}$
is defined by comparing~$\tilde{\mu}$ asymptotically near infinity
with a measure~$\mu$ describing the vacuum.
To this end, one exhausts the supports~$N$ and~$\tilde{N}$ of these measures by sequences of compact
subsets~$(\Omega_n)_{n \in \N}$ and~$(\tilde{\Omega}_n)_{n \in \N}$, respectively,
and takes the limit of a nonlinear surface layer integral of structure similar to~\eqref{osinl},
\begin{align}
&\Mass(\tilde{\mu}, \mu) = \lim_{\Omega_n \nearrow N, \; \tilde{\Omega}_n \nearrow \tilde{N}
\text{ with } \mu(\Omega_n) = \tilde{\mu}(\tilde{\Omega}_n) < \infty} \notag \\
&\quad \times \bigg(
\int_{\tilde{\Omega}_n} \!\!d\tilde{\mu}(\x) \int_{N\setminus \Omega_n} \!\!\!\!\!\!\!\!d\mu(\y)\: \L_\text{static}(\x,\y)
- \int_{\Omega_n} \!\!d\mu(\x) \int_{\tilde{N} \setminus \tilde{\Omega}_n}  \!\!\!\!\!\!\!\!d\tilde{\mu}(\y)\: \L_\text{static}(\x,\y) \bigg) \:.
\label{massintro}
\end{align}
In order for this expression to be well-defined, one needs to assume that the spacetimes are
{\em{asymptotically flat}} (for brevity, we do not give the precise definition, which can be found
in~\cite[Definition~1.5]{pmt}). As a consequence of the EL equations, the mass does not
depend on the choice of the exhaustions, except that the volume condition~$\mu(\Omega_n) = \tilde{\mu}(\tilde{\Omega}_n)$ must hold.
Moreover, a {\em{positive mass theorem}} is proved which states that the total mass~$\Mass(\tilde{\mu}, \mu)$
is non-negative if a suitable {\em{local energy condition}} holds
(for details see~\cite[Definition~1.8 and Theorem~1.9]{pmt}).

The correspondence to the classical positive mass theorem 
(see for example~\cite{schoen+yau}) is established by showing
that for a static causal fermion systems describing the Schwarzschild spacetime
(constructed again as explained in Section~\ref{seccfslorentz}), the total mass~$\Mass(\tilde{\mu}, \mu)$
coincides (up to an irrelevant prefactor) with the ADM mass~\cite{adm} (see~\cite[Theorem~1.10]{pmt}).
For brevity, we cannot enter the proof and the detailed constructions. But we make one remark
which will be important in connection with the construction of the quantum state in Section~\ref{secQFT}:
When comparing causal fermion systems describing the Schwarzschild spacetime and Minkowski space,
one needs to take into account that the resulting causal fermion systems are defined on two different
Hilbert spaces, namely a Hilbert space~$\tilde{\H}$ formed of Dirac solutions in Schwarzschild
and a Hilbert space~$\H$ formed of Dirac solutions in Minkowski space.
Before we can make sense of the nonlinear surface layer integral~\eqref{massintro},
these Hilbert spaces must be identified by a unitary transformation~$V$,
\beq \label{Videntify}
V \::\: \H \rightarrow \tilde{\H} \:.
\eeq
This identification is {\em{not}} canonical but leaves us with the freedom of choosing a unitary transformation.
Therefore, in order to make sense of~\eqref{massintro}, one must prove that the total mass
is independent of the choice of the unitary transformation~$V$ (for details see~\cite[Section~4.3]{pmt}).

\subsection{Connection Between Area Change and Matter Flux} \label{secjacobson}
In 1995, Ted Jacobson gave a derivation of the Einstein equations from thermodynamic principles~\cite{jacobsonarea}.
At the heart of his argument is the formula
\beq \label{am}
\frac{d}{d\tau} A(S_\tau) = c\, F(S_\tau)
\eeq
which states that the area change of a family of two-surfaces~$S_\tau$ propagating along a null Killing direction
is proportional to the matter flux~$F(S_\tau)$ across these surfaces
(with~$c$ a universal constant). In~\cite{jacobson}, this formula is derived in the setting of causal fermion
systems from the EL equations~\eqref{EL} (without referring to thermodynamics).
This result gives an alternative derivation of the Einstein equations from the causal action principle
(again with an undetermined cosmological constant).
Compared to the continuum limit analysis described in Section~\ref{seccl},
this alternative derivation has the advantage that it is more conceptual and thus
gives a more direct understanding of why and how the Einstein equations arise. Moreover,
the procedure in~\cite{jacobson} has the benefit that it does not rely on tensor calculus.
Therefore, it goes beyond Lorentzian geometry and applies to more general ``quantum'' spacetimes.

\begin{figure}
\psscalebox{1.0 1.0} 
{
\begin{pspicture}(0,27.787453)(13.665548,30.652548)
\definecolor{colour1}{rgb}{0.6,0.6,0.6}
\definecolor{colour0}{rgb}{0.8,0.8,0.8}
\definecolor{colour2}{rgb}{0.4,0.4,0.4}
\pspolygon[linecolor=colour1, linewidth=0.02, fillstyle=solid,fillcolor=colour1](0.020444335,28.987547)(0.64044434,29.017548)(1.2304443,29.077547)(1.8804443,29.117548)(2.5804443,29.177547)(3.1904442,29.147547)(3.1904442,28.927547)(3.1604443,28.527548)(3.1454444,28.167547)(3.1954443,27.837547)(0.025444336,27.807549)
\pspolygon[linecolor=colour0, linewidth=0.02, fillstyle=solid,fillcolor=colour0](3.1554444,30.607548)(6.425444,30.617548)(6.425444,28.777548)(5.4854445,28.867548)(4.3754444,29.067547)(3.8854444,29.127548)(3.5054443,29.157547)(3.2354443,29.177547)(3.2654443,29.617548)(3.2254443,30.107548)
\pspolygon[linecolor=colour0, linewidth=0.02, fillstyle=solid,fillcolor=colour0](9.114016,27.892548)(10.298302,27.887548)(10.329016,28.442547)(10.431159,29.107548)(10.38473,29.682549)(10.284016,30.122547)(10.214016,30.577547)(10.230444,30.632547)(8.830444,30.632547)(7.2304444,30.632547)(7.23973,30.002548)(7.226873,29.222548)(7.2254443,28.677547)(7.229016,28.167547)(7.2254443,27.887548)
\pspolygon[linecolor=colour1, linewidth=0.04, fillstyle=solid,fillcolor=colour1](7.2254443,28.982548)(7.820444,28.982548)(8.075444,29.007547)(8.510445,29.057549)(8.915444,29.097548)(9.650444,29.142548)(10.270444,29.147547)(11.000444,29.097548)(11.745444,29.007547)(12.285444,28.912548)(12.880445,28.812548)(13.4204445,28.767548)(13.635445,28.757547)(13.630445,27.832548)(7.2304444,27.832548)
\psbezier[linecolor=colour2, linewidth=0.04](10.230444,30.632547)(10.320444,29.622547)(10.548764,29.309046)(10.400444,28.73743696464)(10.252125,28.165827)(10.29298,27.994743)(10.280444,27.817547)
\rput[bl](7.3604445,30.177547){\normalsize{$V$}}
\psbezier[linecolor=black, linewidth=0.04](7.2104445,28.997547)(8.211427,28.975313)(9.091899,29.225868)(10.532664,29.157547607421876)(11.973431,29.089228)(12.276996,28.840084)(13.645444,28.767548)
\rput[bl](10.990444,28.037548){\normalsize{$\Omega$}}
\pscircle[linecolor=black, linewidth=0.04, fillstyle=solid,fillcolor=black, dimen=outer](10.430445,29.162548){0.095}
\rput[bl](10.555445,29.277548){\normalsize{$S$}}
\psline[linecolor=black, linewidth=0.02, fillstyle=solid,fillcolor=black, arrowsize=0.05291667cm 2.0,arrowlength=1.4,arrowinset=0.0]{->}(10.695444,29.847548)(10.630445,30.267548)
\rput[bl](9.5454445,29.622547){\normalsize{$v$}}
\psline[linecolor=black, linewidth=0.02, fillstyle=solid,fillcolor=black, arrowsize=0.05291667cm 2.0,arrowlength=1.4,arrowinset=0.0]{->}(9.335444,29.137548)(9.230444,29.537548)
\psline[linecolor=black, linewidth=0.02, fillstyle=solid,fillcolor=black, arrowsize=0.05291667cm 2.0,arrowlength=1.4,arrowinset=0.0]{->}(8.470445,29.067547)(8.390445,29.392548)
\psline[linecolor=black, linewidth=0.02, fillstyle=solid,fillcolor=black, arrowsize=0.05291667cm 2.0,arrowlength=1.4,arrowinset=0.0]{->}(7.720444,29.002548)(7.7154446,29.352547)
\psline[linecolor=black, linewidth=0.02, fillstyle=solid,fillcolor=black, arrowsize=0.05291667cm 2.0,arrowlength=1.4,arrowinset=0.0]{->}(9.980444,29.792547)(9.910444,30.197548)
\psline[linecolor=black, linewidth=0.02, fillstyle=solid,fillcolor=black, arrowsize=0.05291667cm 2.0,arrowlength=1.4,arrowinset=0.0]{->}(9.205444,29.747547)(9.120444,30.227547)
\psline[linecolor=black, linewidth=0.02, fillstyle=solid,fillcolor=black, arrowsize=0.05291667cm 2.0,arrowlength=1.4,arrowinset=0.0]{->}(10.060444,29.187548)(10.0454445,29.667547)
\psline[linecolor=black, linewidth=0.02, fillstyle=solid,fillcolor=black, arrowsize=0.05291667cm 2.0,arrowlength=1.4,arrowinset=0.0]{->}(7.6404443,29.497547)(7.5904446,29.992548)
\psline[linecolor=black, linewidth=0.02, fillstyle=solid,fillcolor=black, arrowsize=0.05291667cm 2.0,arrowlength=1.4,arrowinset=0.0]{->}(8.340445,29.547548)(8.255445,30.027548)
\psline[linecolor=black, linewidth=0.02, fillstyle=solid,fillcolor=black, arrowsize=0.05291667cm 2.0,arrowlength=1.4,arrowinset=0.0]{->}(11.080444,29.137548)(11.075444,29.557549)
\psline[linecolor=black, linewidth=0.02, fillstyle=solid,fillcolor=black, arrowsize=0.05291667cm 2.0,arrowlength=1.4,arrowinset=0.0]{->}(11.745444,29.052547)(11.735444,29.482548)
\psline[linecolor=black, linewidth=0.02, fillstyle=solid,fillcolor=black, arrowsize=0.05291667cm 2.0,arrowlength=1.4,arrowinset=0.0]{->}(12.300445,28.937548)(12.275444,29.352547)
\psline[linecolor=black, linewidth=0.02, fillstyle=solid,fillcolor=black, arrowsize=0.05291667cm 2.0,arrowlength=1.4,arrowinset=0.0]{->}(10.015445,28.327547)(10.070444,28.732548)
\psline[linecolor=black, linewidth=0.02, fillstyle=solid,fillcolor=black, arrowsize=0.05291667cm 2.0,arrowlength=1.4,arrowinset=0.0]{->}(10.585444,28.347548)(10.640445,28.797548)
\psbezier[linecolor=colour2, linewidth=0.04](3.2004442,27.817547)(3.1204443,28.332548)(3.1621246,28.510939)(3.2304444,29.232547607421875)(3.298764,29.954157)(3.1379087,30.45535)(3.1504443,30.632547)
\psbezier[linecolor=black, linewidth=0.04](0.010444336,28.997547)(1.0114269,28.975313)(1.8918986,29.225868)(3.3326645,29.157547607421876)(4.7734303,29.089228)(5.076996,28.840084)(6.445444,28.767548)
\pscircle[linecolor=black, linewidth=0.04, fillstyle=solid,fillcolor=black, dimen=outer](3.2304444,29.162548){0.095}
\rput[bl](3.3754444,29.317547){\normalsize{$S$}}
\rput[bl](1,28.3){\normalsize{$\Omega \cap V$}}
\rput[bl](4,29.7){\normalsize{$M \setminus (\Omega \cup V)$}}
\end{pspicture}
}
\caption{Two-dimensional surface layer integrals.}
\label{figarea}
\end{figure}
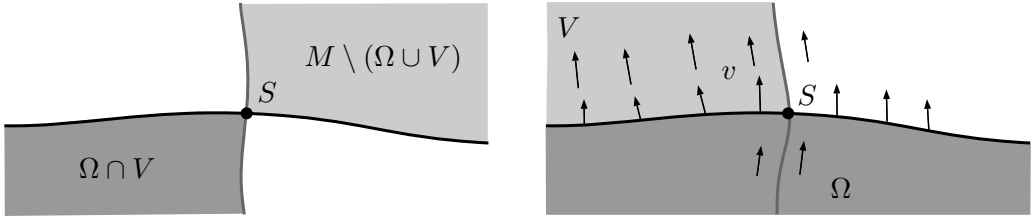

We now explain the general procedure and a few constructions from~\cite{jacobson}.
The first question is how to define the area~$A$ of a two-dimensional surface~$S$ in the setting of causal
fermion systems. It is most convenient to describe~$S \subset M$ as
\[ S = \partial \Omega \cap \partial V \:, \]
where~$\Omega$ can be thought of as being the past of a Cauchy surface, and~$V$ describing
a spacetime cylinder. This description has the advantage that the resulting surface layer integrals
are well-defined even in situations when spacetime is singular or discrete, in which
case the boundaries~$\partial \Omega$ and~$\partial V$ are no longer a sensible concept.
The most natural way of introducing a surface layer integral localized in a neighborhood of~$S$
is a double integral of the form
\beq \label{osi2d1}
\int_{\Omega \cap V} \bigg( \int_{M \setminus (\Omega \cup V)} (\cdots)\: \L_\kappa(x,y)\: d\rho(y) \bigg)\, d\rho(x)
\eeq
(where~$(\cdots)$ stands again for a differential operator acting on the Lagrangian).
If the Lagrangian has short range, we only get contributions to this surface layer integral if both~$x$
and~$y$ are close to the two-dimensional surface~$S$ (see the left of Figure~\ref{figarea}).

The disadvantage of this method is that, similar as explained in Section~\ref{seclqg}
for the structures of the Lorentzian quantum geometry and the EL equations,
the surface layer integral~\eqref{osi2d1} does not fit together with the notion of matter flux.
Therefore, it is preferable to define the two-dimensional area as follows.
We need to assume that~$M$ has a smooth manifold structure, and that the measure~$\rho$
is absolutely continuous and smooth with respect to the Lebesgue measure in every chart~$(x,U)$ of~$M$, i.e.\
\beq \label{rhoh}
d\rho = h(x)\: d^4x \qquad \text{with~$h \in C^\infty(U, \R^+)$}\:.
\eeq
Also assuming that the boundaries of~$\Omega$ and~$V$ are smooth, we can choose
a smooth vector field~$v$ which is transverse to the
hypersurface~$\partial \Omega$ and tangential to~$\partial V$ (see the right of Figure~\ref{figarea}).
Under these additional assumptions, one can introduce an integration measure~$d\mu$
on the hypersurface~$\partial \Omega$ by
\[ 
d\mu(v,x) = h\: \epsilon_{ijkl} \:v^i\: dx^j dx^k dx^l \:, \]
where~$\epsilon_{ijkl}$ is the totally anti-symmetric Levi-Civita symbol
(normalized by~$\epsilon_{0123}=1$).
Now we can define the two-dimensional area by
\beq \label{Adef}
A := \int_{\partial \Omega \cap V} d\mu(v, x) \int_{M \setminus V} d\rho(y)\: \L_\kappa(x,y) \:.
\eeq

\begin{figure}
%
\psscalebox{1.0 1.0} 
{
\begin{pspicture}(-3.5,-1.2303988)(10.170529,1.2303988)
\definecolor{colour0}{rgb}{0.8,0.8,0.8}
\definecolor{colour1}{rgb}{0.6,0.6,0.6}
\definecolor{colour2}{rgb}{0.4,0.4,0.4}
\pspolygon[linecolor=colour0, linewidth=0.02, fillstyle=solid,fillcolor=colour0](1.9141006,-1.2053374)(3.0983863,-1.2103374)(3.0241005,-0.6453374)(3.0312436,0.0)(3.0148149,0.5596626)(3.0091007,0.7396626)(3.0041006,0.9946626)(3.0055292,1.2196625)(1.5091006,1.2096626)(0.032672033,1.1946626)(0.03981489,0.9046626)(0.026957747,0.12466259)(0.025529174,-0.4203374)(0.029100602,-0.9303374)(0.025529174,-1.2103374)
\pspolygon[linecolor=colour1, linewidth=0.04, fillstyle=solid,fillcolor=colour1](0.020529175,-0.4153374)(0.6155292,-0.4153374)(0.8705292,-0.3903374)(1.3055291,-0.3403374)(1.7105292,-0.3003374)(2.4455292,-0.25533742)(3.065529,-0.25033742)(3.7955291,-0.3003374)(4.5405293,-0.3903374)(5.080529,-0.4853374)(5.675529,-0.5853374)(6.215529,-0.6303374)(6.430529,-0.6403374)(6.430529,-1.2103374)(0.040529177,-1.2003374)
\psbezier[linecolor=colour2, linewidth=0.04](3.0105293,1.2296625)(2.994465,0.7911212)(3.0596163,-0.02383872)(3.0405293,-0.360448052755018)(3.021442,-0.69705737)(3.0112207,-1.0331409)(3.007802,-1.2103374)
\rput[bl](0.18552917,0.6046626){\normalsize{$V$}}
\psbezier[linecolor=black, linewidth=0.04](0.0105291745,-0.4003374)(1.0115118,-0.4225725)(1.8919835,-0.17201768)(3.3327494,-0.24033740997314454)(4.773515,-0.30865714)(5.0770807,-0.5578018)(6.445529,-0.6303374)
\rput[bl](5.990529,-1.0603374){\normalsize{$\Omega$}}
\pscircle[linecolor=black, linewidth=0.04, fillstyle=solid,fillcolor=black, dimen=outer](3.0405293,-0.2203374){0.095}
\rput[bl](3.1555293,0.04966259){\normalsize{$S$}}
\psline[linecolor=black, linewidth=0.02, arrowsize=0.05291667cm 2.0,arrowlength=1.4,arrowinset=0.0]{->}(2.6955292,-0.22533742)(2.6555293,0.21966259)
\rput[bl](2.4105291,0.5646626){\normalsize{$v$}}
\psline[linecolor=black, linewidth=0.02, arrowsize=0.05291667cm 2.0,arrowlength=1.4,arrowinset=0.0]{->}(2.1355293,-0.2603374)(2.105529,0.14466259)
\psline[linecolor=black, linewidth=0.02, arrowsize=0.05291667cm 2.0,arrowlength=1.4,arrowinset=0.0]{->}(1.2705292,-0.3303374)(1.2455292,0.06966259)
\psline[linecolor=black, linewidth=0.02, arrowsize=0.05291667cm 2.0,arrowlength=1.4,arrowinset=0.0]{->}(0.52052915,-0.3953374)(0.49552917,0.00966259)
\psline[linecolor=black, linewidth=0.02, arrowsize=0.05291667cm 2.0,arrowlength=1.4,arrowinset=0.0]{->}(2.7805293,0.3946626)(2.740529,0.8746626)
\psline[linecolor=black, linewidth=0.02, arrowsize=0.05291667cm 2.0,arrowlength=1.4,arrowinset=0.0]{->}(2.0055292,0.3496626)(1.9555292,0.7746626)
\psline[linecolor=black, linewidth=0.02, arrowsize=0.05291667cm 2.0,arrowlength=1.4,arrowinset=0.0]{->}(3.0455291,-0.18533741)(3.0305293,0.2946626)
\psline[linecolor=black, linewidth=0.02, arrowsize=0.05291667cm 2.0,arrowlength=1.4,arrowinset=0.0]{->}(3.0053284,-1.017598)(3.59073,-1.0230768)
\psline[linecolor=black, linewidth=0.02, arrowsize=0.05291667cm 2.0,arrowlength=1.4,arrowinset=0.0]{->}(0.87552917,0.34466258)(0.83552915,0.8146626)
\psline[linecolor=black, linewidth=0.02, arrowsize=0.05291667cm 2.0,arrowlength=1.4,arrowinset=0.0]{->}(3.0253847,-0.58398014)(3.6106737,-0.59669465)
\psline[linecolor=black, linewidth=0.02, arrowsize=0.05291667cm 2.0,arrowlength=1.4,arrowinset=0.0]{->}(3.0804245,-0.2433218)(3.665634,-0.227353)
\rput[bl](3.690529,-0.6503374){\normalsize{$u$}}
\end{pspicture}
}
\caption{Matter flux through~$S$.}
\label{figflux}
\end{figure}
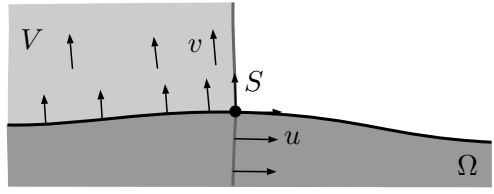

We next describe a symmetry by a notion of a Killing field. This notion will also
give rise to a notion of matter flux.
\begin{Def} \label{defkilling}
A vector field~$u$ on~$M$ is called {\bf{Killing field}} of the causal fermion system if
the following conditions hold:
\bitem
\item[\rm{(i)}] The divergence of~$u$ vanishes, i.e.
\[ \frac{1}{h}\: \partial_j \big( h\, u^j \big) = 0 \]
(where~$h$ is again the weight function in~\eqref{rhoh}).
\item[\rm{(ii)}] The directional derivative of the Lagrangian is small in the sense that
\beq \label{DuL}
\big( D_{1,u} + D_{2,u} \big) \L_\kappa(x,y) \lesssim \frac{m^4}{\varepsilon^4\: \delta^4}
\eeq
(where again~$\varepsilon$ is the regularization length, and~$\delta$ is the Planck length).
\eitem
\end{Def}
The reason why the right side in~\eqref{DuL} is non-zero is that the derivative of the Lagrangian
involves contributions from both the geometry and the matter fields.
The condition~\eqref{DuL} means that the geometric contributions vanish
(which are typically much larger). The fact that the matter contribution remains
makes it possible to define the matter flux by
\beq \label{FOdef}
F(S) := \int_{\partial \Omega \cap V} d\mu(v, x) \int_{M \setminus V} d\rho(y) \:
\big(D_{1,u} + D_{2,u} \big) \L_\kappa(x,y) \:,
\eeq
where we assume that the Killing field~$u$ is tangential to~$\Omega$ (see Figure~\ref{figflux}).
In the limiting case of lightlike propagation, the vector fields~$u$ and~$v$ coincide,
making it possible to recover the matter flux~\eqref{FOdef} as the derivative of
the area~\eqref{Adef} in the direction of the Killing field~$u$. This gives the
desired relation~\eqref{am}.

\subsection{The Quantum State of a Causal Fermion System} \label{secQFT}
We finally give an outlook on quantum field theory and quantum gravity.
So far, the connection to quantum field theory has been established only
for causal fermion systems describing Minkowski space~\cite{fockbosonic, fockfermionic}.
But most of the methods apply just as well to more general causal fermion systems describing curved spacetimes.
Moreover, one should keep in mind that in a causal fermion system, all the bosonic interactions
are described in a unified way with the same mathematical structures, giving hope that methods
developed for the electromagnetic interaction in Minkowski space can be adapted to the gravitational field.
But, as is usually the case in mathematics, the connections and analogies are rather subtle and need to be
explored carefully and in depth. We plan to do so in the near future.

Coming back to causal fermion systems describing Minkowski space, 
in~\cite{fockfermionic} a {\em{quantum state}}~$\omega^t$ of a causal fermion system~$(\tilde{\H}, \tilde{\F}, \tilde{\rho})$, which describes an interacting spacetime, is constructed.
Here by a ``quantum state'' we mean a positive linear functional on the $*$-algebra of
observables~$\A$, i.e.
\[ \omega^t \::\: \A \rightarrow \C \quad \text{complex linear} \qquad \text{and} \qquad
\omega^t \big( A^* A \big) \geq 0 \quad \text{for all~$A \in \A$}\:. \]
The observables are formed of linearized solutions of the EL equations.
The general idea for constructing the quantum state is to compare~$(\tilde{\H}, \tilde{\F}, \rho)$ at time~$t$
with a causal fermion system~$(\H, \F, \rho)$ describing the vacuum.
Similar to the procedure for the total mass~\eqref{massintro}, the ``comparison'' is again
performed with a nonlinear surface layer integral~\eqref{osinl}, but now with~$\tilde{\Omega}$ and~$\Omega$
chosen as the past of the respective Cauchy surfaces (similar as shown in Figure~\ref{figosinl}).
The unitary freedom in identifying the Hilbert spaces already mentioned in the connection of the
positive mass theorem in~\eqref{Videntify} is now taken care of by integrating over a group~$\G$
of unitary operators.
This leads to a formula for the quantum state of the symbolic form
\beq \label{statedef}
\omega^t( A) := \frac{1}{Z^t\big( \beta, \tilde{\rho} \big)} \:
\int_\G (\cdots)\: e^{\beta\, \gamma^t(\tilde{\rho}, \scrU \rho)} \; d\mu_\G(\scrU) \:,
\eeq
where~$Z^t\big( \beta, \tilde{\rho} \big)$ is the partition function defined by
\[ Z^t\big( \beta, \tilde{\rho} \big) :=
\int_\G e^{\beta\, \gamma^t(\tilde{\rho}, \scrU \rho)} \;
d\mu_\G(\scrU) \:, \]
and~$\beta$ is a real parameter. The factor~$(\cdots)$ in~\eqref{statedef} consists
of a product of surface layer integrals formed of the linearized solutions contained in the observable~$A$
(for details see~\cite[Section~4]{fockfermionic}).

The significance of this construction is that it becomes possible to describe the interacting
causal fermion system in the familiar language of quantum field theory.
Consequently, also the dynamics as described by the EL equations~\eqref{EL} 
can be rewritten as a time evolution of the state~$\omega^t$. The detailed form of
the resulting quantum dynamics is presently under investigation~\cite{mix}.

\Thanks{{{\em{Acknowledgments:}} I would like to thank Michael Kiessling and
Shadi Tahvildar-Zadeh for the kind invitation to the Marcel Grossmann Meeting.
I am grateful to Christoph Krpoun for helpful comments on the manuscript.

\providecommand{\bysame}{\leavevmode\hbox to3em{\hrulefill}\thinspace}
\providecommand{\MR}{\relax\ifhmode\unskip\space\fi MR }
\providecommand{\MRhref}[2]{%
  \href{http://www.ams.org/mathscinet-getitem?mr=#1}{#2}
}
\providecommand{\href}[2]{#2}

\end{document}